# Thermal expansion of organic superconductor α-(BEDT-TTF)$_2$ NH$_4$Hg(SCN)$_4$.

A.V. Dolbin[1], M.V. Khlistuck[1], V.B. Eselson[1], V.G. Gavrilko[1], N.A. Vinnikov[1],
R.M. Basnukaeva[1], V.A. Konstantinov[1], K.R. Luchinskii[2], Y. Nakazawa[3]

[1] *B. Verkin Institute for Low Temperature Physics and Engineering, NAS of Ukraine, 47 Nauky Ave., Kharkov, 61103, Ukraine.*

[2] *V.N. Karazin Kharkiv National University, 4 Svobody Sq., Kharkov, 61022, Ukraine*

[3] *Department of Chemistry, Graduate School of Science, Osaka University, 1-1 Machikaneyama-cho, Toyonaka, Osaka 560-0043, Japan*

e-mail:  dolbin@ilt.kharkov.ua

**Abstract.**

The temperature dependence of the linear thermal expansion coefficients (LTEC) of a single crystal of *α*-(BEDT-TTF)$_2$NH$_4$Hg(SCN)$_4$ where BEDT-TTF is bis(ethylenedithio)tetrathiafulvalene was studied by the method of precision capacitive dilatometry in the temperature range 2-250 K along the crystallographic direction *b\** (perpendicular to the crystal layers). Negative values of LTEC were found below 4 K. Probably it is due to charge fluctuations as the temperature approaches the temperature of transition to the superconducting state $T_c$. It has been suggested that the bends on temperature dependence of LTEC observed in the temperature range 25-45 K are related to order-disorder arrangement of NH$_4^+$ ions. Weak maximum of the LTEC, detected at a temperature of about 200-220 K, can be caused by the processes of charge redistribution and the associated intermolecular interaction fluctuations.

**Introduction.**
Organic (super) conductors are a series of multi-components molecular compounds categorized as organic charge transfer salts. They have a variety of electronic phases produced by $\pi$- electrons/holes released from molecular orbitals of organic molecules (HOMOs/LUMOs). They usually form an alternatively stacked structure of organic molecule layers and counter-ion layer. The molecular arrangements in the organic molecule layer are closely related to the band structures determined by the various electron transfers between neighboring molecules in the conducting layers. Electronic state at low temperatures shows divers variations due to the difference of these structural features. The itinerant $\pi$–carriers also have feature of electron correlations as well as the characters induced by strong electron-lattice coupling. The development of these molecule-based conducting compounds was initiated by the ideas of Little and Ginzburg [1, 2] on the possibility of superconductivity in quasi-one-dimensional and two-dimensional heterogeneous structures in which layers with metallic and dielectric properties alternate. The well-known component molecule of most numerous organic (super) conductors is the organic molecule BEDT-TTF (bis(ethylenedithio)tetrathiafulvalene, also often abbreviated as ET). The BEDT-TTF molecule consists of two five-membered and two six-membered rings, each of which contains two sulfur atoms [3]. These molecules are almost planar, which makes it possible to pack them in the form of quasi-one dimensional stacking, or uniform or zig-zag stackings of dimerized units, etc. (Fig. 1).

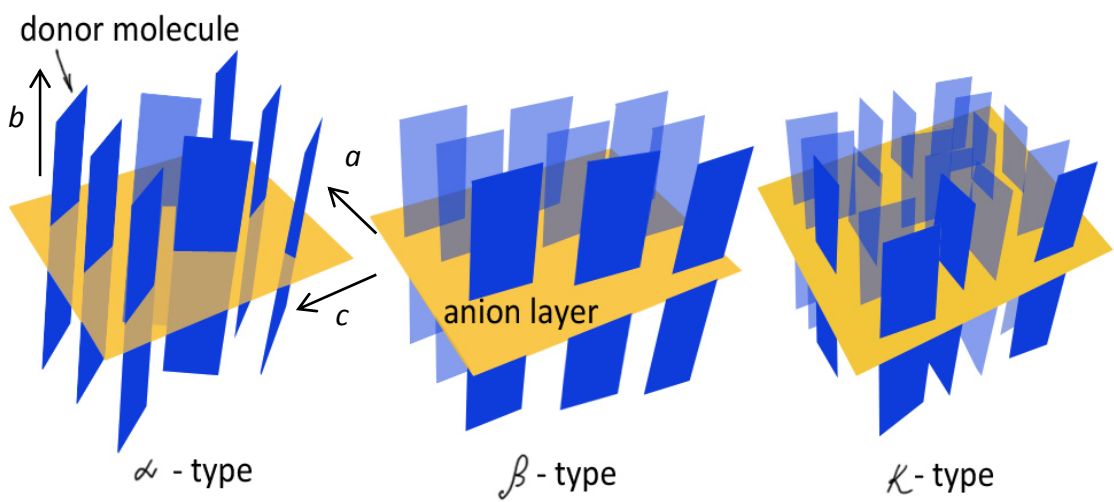

Fig. 1. Location of molecular layers in organic conductors of *α*, *β* and *κ*-type.

The unpaired $\pi$-electrons of the HOMO delocalized over carbon and sulfur atoms of the BEDT-TTF molecule form a set of $\pi$-orbitals perpendicular to the plane of molecule [4, 5]. The ET molecules are donors of a charge (one electron per two ET molecules [6]) and form ordered conductive layers in the crystal, separated by acceptor insulating layers (in our case, by the [NH$_4$Hg(SCN)$_4$]$^-$ anions).

Organic conductors of the family $\alpha$ - (BEDT-TTF)$_2$MHg(SCN)$_4$ (where M$^+$ = NH$_4^+$, K$^+$, Rb$^+$, Tl$^+$) are characterized as metallic salts with quasi-1D and 2D Fermi surface. Organic conductors $\alpha$-(ET)$_2$KHg(SCN)$_4$ and $\alpha$-(ET)$_2$NH$_4$Hg(SCN)$_4$ were first synthesized as possible superconducting modifications of the compound $\kappa$-(ET)$_2$Cu(SCN)$_2$ possessing a 10 K transition temperature [7]. Despite of the fact that these two salts have isostructure categorized as non-dimerized α-type packing their low temperatures properties are very different. $\alpha$-(ET)$_2$NH$_4$Hg(SCN)$_4$ is a superconductor with $T_c$ ~1.15 K, whereas $\alpha$-(ET)$_2$KHg(SCN)$_4$ has a metallic conductivity down to 100 mK [8,9]. Kondo et al., based on combined discussion by transport properties under uniaxial stress measurements and band structure calculations [10], suggested that the quasi- one-dimensional *Q1D* zone contributes to superconductivity [11]. The application of external pressure gives two effects on $\alpha$-(ET)$_2$NH$_4$Hg(SCN)$_4$: first, when hydrostatic pressure is applied, superconductivity disappears, and secondly, at *P* larger than 0.2 GPa the band structure of this compound is modified, which leads to peculiarities of magnetic and transport properties in the non superconducting phase [12]. It was also noted that the presence in the structure of compound $\alpha$-(BEDT-TTF)$_2$(M)Hg(SCN)$_4$ groups NH$_4^+$ (in contrast to monatomic *M* metal ions) introduce an orientational freedom that may induce disorder into the crystal, which leads to disordering of open π-electron orbits [12]. This ensures the possibility of superconductivity at low temperatures and normal pressure in $\alpha$-(BEDT-TTF)$_2$NH$_4$Hg(SCN)$_4$. This assumption is confirmed by the relationship between the parameters *a* and *c* of the elementary crystal cell of this compound [13]. For compounds containing NH$_4^+$ ions, a decrease in the lattice parameter ratio *c/a* enhances superconductivity, while an increase in *c/a* stabilizes the metallic state. Intermediate between them is the density-wave state. The characteristic changes in the electronic properties of these compounds can be systematically represented as a function of *c/a*. Due to anisotropic thermal compression, the ratio *c/a* for the NH$_4$ salt is much smaller than for M = K, Rb, Tl at low temperatures and normal pressure, which explains the transition of NH$_4^+$ compounds to the superconducting state. In these compounds, the phase transition to a charge-ordered state is due to the competition between the kinetic energy of the system, and the Coulomb energy (external and internal) of nearest neighbors, ET molecules [14]. Based on the phase diagram [15], it can be concluded that as the electronic correlations increase, the ground state of these systems changes from the metallic (Fermi liquid) to the Mott insulator, in addition, there is a superconductivity region between these states [13, 16].

Earlier we investigated using a low-temperature capacitance dilatometer the low-temperature thermal expansion of quasi-two-dimensional organic conductors

κ-(BEDT-TTF)$_2$Cu[N(CN)$_2$]Cl and κ-(D$_4$-BEDT-TTF)$_2$Cu[N(CN)$_2$]Br along the crystallographic axes *a* and *b* in the temperature range 2-290 K [19,20]. The temperature dependences of the linear thermal expansion coefficients (LTEC) α(T) of these single crystals were determined and a correlation between the anomalies in their thermal expansion and the phase transformations was established. The purpose of this work was to investigate the influence of structural features and the phase state of α-(BEDT-TTF)$_2$NH$_4$Hg(SCN)$_4$ compound on the thermal expansion of this crystal.

**Experimentals**

Organic salt α-(BEDT-TTF)$_2$NH$_4$Hg(SCN)$_4$ has an orthorhombic two-layered crystal lattice with four ET molecules in the unit cell. The layers of anions in this compound are located in the *ac*-plane and are perpendicular to the *b*-axis along which the flat BEDT-TTF molecules are located [18]. Conductive and insulating layers alternate along the $b^*$ axis (Fig. 1). Positions of the crystallographic axes in the studied sample, the crystal structure and parameters of lattice was confirmed by x-ray diffraction (at room temperature *a* = 10.103 Å, *b* = 20.557 Å, *c* = 9.935 Å).

The study of linear thermal expansion coefficient of α-(BEDT-TTF)$_2$NH$_4$Hg(SCN)$_4$ single crystal was performed in the direction of the crystallographic axis *b* perpendicular to the crystal layers in the temperature range 2-250 K using a low-temperature capacitive highly sensitive dilatometer [19]. Since this direction is the inter-layer direction, the thermal conductivity seems sensitive to the change of the order-disorder arrangement of the anion layer. The change of this direction is also related to the 2D features in the transport and magnetic properties of this compound. The length of the sample in the measuring direction was *l* = 0.7 mm. The measured values of the coefficient of linear thermal expansion are shown in Fig. 2.

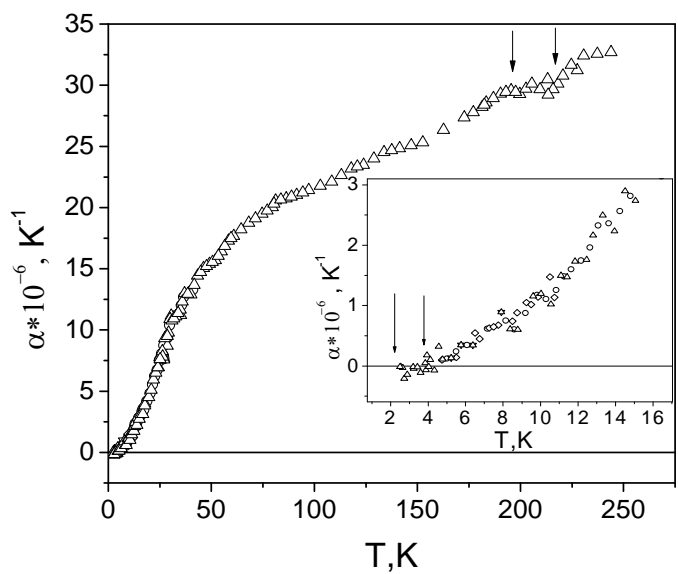

a)

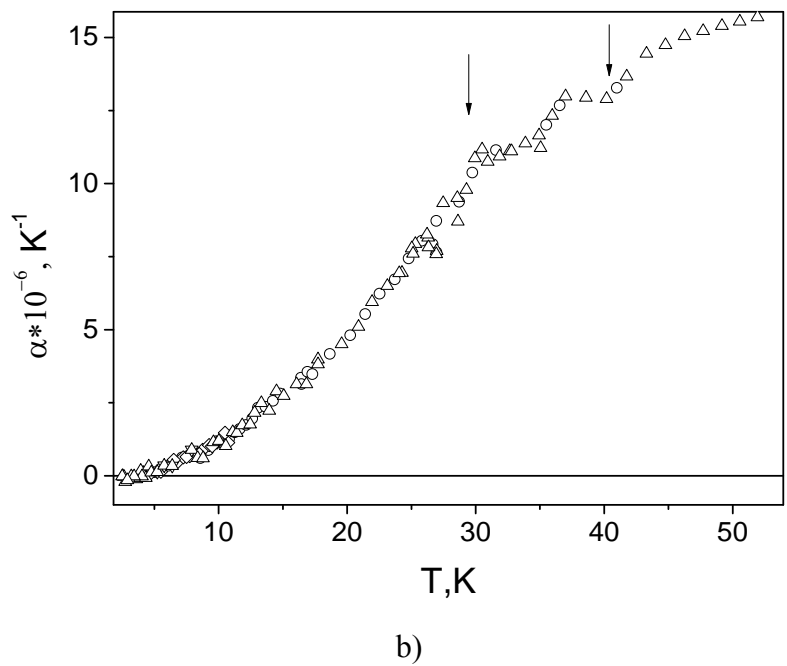

b)

Fig.2. The dependence of LTEC on temperature a) 2-250 K; b) 2-55 K. The interval 2-17 K is also shown in the inset to Fig. 2a. Different symbols correspond to data from different series of experiments. Arrows indicate the temperature intervals of LTEC anomalies.

**Results and discussion.**

Temperature dependence of LTEC of the single crystal sample of α-(BEDT-TTF)$_2$NH$_4$Hg(SCN)$_4$ do not have pronounced peaks observed earlier in the LTEC dependences of single crystals κ-(D$_4$-BEDT-TTF)$_2$Cu[N(CN)$_2$]Br and κ-(BEDT-TTF)$_2$Cu[N(CN)$_2$]Cl [19,20]. It decreases almost linear from room temperature down to about 100K as is shown in the Fig. 2a). The decrease become steep below about 70K where the ethylene groups of BEDT-TTF molecules seem to be frozen or ordered. Since we could not detect peak like structure related to the ethylene dynamics in κ-type compounds, the lattice confliction seems to be quite normal for molecular compounds. However, there are a number of features of $α(T)$ (see Fig. 2) which must be related to the electron-lattice coupling of this compound.

Below 3.5-4 K LTEC of α-(BEDT-TTF)$_2$NH$_4$Hg(SCN)$_4$ has small negative values (see the insert to Fig. 2a). It is assumed that the negative thermal expansion coefficient of this compound can be due to formation of hindered tunneling level of NH$_4^+$ or superconductive fluctuations coupled with charge disproportionation [15]. In the case of κ-type compounds, the negative thermal expansion is observed only below $T_c$. It is interesting to note that the attribute of charge fluctuations in the family of α-(BEDT-TTF)$_2$MHg(SCN)$_4$ (where M – NH$_4$, Rb, Tl) most clearly manifest itself in NH4 compound, since in ET-NH4 salt the isolating anionic layer is thick among various of ET-based superconductor molecules. The period of the layer in the direction perpendicular to the plane is about 20 A°. The heat capacity measurements of those compounds performed previously demonstrated that the fluctuation region of superconductivity ranges to the higher temperature region and hindered rotation contribution of heat capacity is observed as upturn of $C_p T^{-1}$ [21, 22]. These thermodynamic features should be related to the appearance of negative thermal expansion region at low temperatures.

There are several anomalies (bends) of the dependence $\alpha(T)$ (Fig. 2b) in temperature range 25-45 K. The magnitude of these anomalies exceeds the experimental error. In this temperature range may be a kind of order-disorder arrangements of $NH_4^+$. The preliminary thermal conductivity measurement of the dueterated compound ($ND_4^+$) gives an anomaly around 25 K [23]. Probably the $NH_4^+$ containing compound show it at higher temperature than the deuterated compound and may show glass-like character, since the thermal expansion shows two-step behavior as is shown in Fig. 2b).

The anomaly in the higher temperature region is considered as an electronic origin peculiar for non-dimeric arrangement. In the case of $\alpha$-type BEDT-TTF compounds, in addition to the local (on-site) Coulomb interaction $U$, the second important parameter that determines the properties of organic salts is the electronic repulsion $V$ between the nearest neighbors. In the case of non-dimerized structure, the metallic state may change to charge ordered insulating states through the increase of $V/W$ ratio. In the region of a quantum critical point, superconductivity can be associated with the possible emergence of charge fluctuations [17]. When cooled, the electronic properties of these systems change due to several factors: (1) as for any metal, the scattering rate of charge carriers decreases, since the phonon oscillations are frozen; (2) a decrease in the volume of the crystalline cell increases proportionally the density of the charge carriers; (3) the optical throughput as well as the electrical conductivity of the crystals increase with increasing transfer integrals $t$ and a decrease in the ratio $V/t$ (the transfer integrals ($t$) are defined as $t = E \cdot S$, where $E$ (= -10 eV) is the empirical constant and $S$ is the overlap integral [18]). Experimentally the M=$NH^{4+}$ compound has the largest two-dimensional feature among $\alpha$-type compounds, since the thick anions layers work to separate the $\pi$-electrons layers [18]. In the temperature range 200-220 K the dependence of $\alpha(T)$ shows a change in the slope and an increased scatter of the experimental values (Fig. 2a). Since this compound do not have dimerized structure, the charge disproportionation and fluctuations due to inter-molecular coulomb repulsions may be important. According to the NMR experiments there is some kind of anomaly related to the charge disproportionation in this temperature region [15]. The hump structure (Fig. 2a) is probably explained by this circumstance since the charge disproportionation itself is electronic origin but should have large change in phonon structure. The change in the angle of inclination of temperature LTEC dependence can be due to freezing of free rotational motion of $NH_4^+$ ions coordinated by four SCN ligands and therefore appears in thermal expansion as a hump structure. [24].

**Conclusions.**

Thermal expansion of a single crystal of α-(BEDT-TTF)$_2$ NH$_4$Hg(SCN)$_4$ was studied by the method of precision capacitive dilatometry in the temperature range 2-250 K along the crystallographic direction *b** axis (perpendicular to the crystal layers). The obtained temperature dependence of LTEC had a number of features due to the phase composition and structure of the compound. The negative LTEC values found below 3-4 K are possibly due to charge fluctuations as the measurement temperature approaches the temperature of transition to the superconducting state $T_c$. The kinks of the *α(T)* dependence in the temperature range 25-45 K are most likely associated with order-disorder arrangement of $NH_4^+$ ions. Weak maximum of the LTEC, detected at temperature of about 200 K, can be caused by the processes of charge redistribution and associated intermolecular interaction fluctuations. The observed change in character of the *α(T)* dependence at 200-220 K is probably correlated with the disinhibition of the rotational motion of $NH_4^+$ ions with increasing temperature.

**Reference**

1. W.A. Little, *Phys. Rev.* **134**, A1416 (1964).
2. V.L. Ginzburg, Sov. Phys. *JETP*, **47**, 2318 (1964).
3. M. Mizuno, A.F. Garito, M.P. Cava, *J. Chem. Soc.,* **1**, 18 (1978).
4. M. Sorai, Y. Nakazawa, M. Nakano and Y. Miyazaki, *Chem. Rev.*, **113**, PR41 (2013).
5. T. Ishiguro, K. Yamaji and G. Saito, *Organic Superconductors*, Berlin, NY: Springer (1996).
6. N. Hanasaki, S. Kagoshima, N. Miura, G. Saito, *Phys. Rev. B.,* **63**, 245116 (2001).
7. J. Singleton, F.L. Pratt, M. Doporto, J. Caulfield and W. Hayes, *Synthetic Metals,* **55-57**, 2198 (1993).
8. F.L. Pratt, J. Singleton, M. Doporto, A.J. Fisher, T.J.B.M. Janssen, J.A.A.J. Perenboom, M. Kurmoo, W. Hayes, and P. Day, *Phys. Rev. B.,* **45**, 13904 (1992).
9. J.S. Brooks, C.C. Agosta, S.J. Klepper, M. Tokumoto, N. Kinoshita, H. Anzai, S. Uji, H. Aoki, A. S. Perel, G. J. Athas, and D. A. Howe, *Phvs. Rev. Lett.,* **69**, 156 (1992).
10. R. Kondo, S. Kagoshima and M. Maesato, *Phys. Rev. B.,* **67**, 134519 (2003).

11. D. Andres, M.V. Kartsovnik, W. Biberacher, K. Neumaier, E. Schuberth, and H. Müller, *Phys. Rev. B.,* **72**, 174513 (2005).
12. S.J. Klepper, J.S. Brooks, X. Chen, I. Bradaric, M. Tokumoto, N. Kinoshita, Y. Tanaka, and C.C. Agosta, *Phys. Rev. B.,* **48**, 9913 (1993).
13. M. Maesato, Y. Kaga, R. Kondo and S. Kagoshima, *Phys. Rev. B.,* **64**, 155104 (2001).
14. K. Yakushi, *Crystals,* **2**, 1291 (2012).
15. Y. Ihara, K. Noda and A. Kawamoto, *Phys. Rev. B.,* **90**, 041107(R) (2014).
16. N. Tajima, A. Ebina-Tajima, M. Tamura, Y. Nishio, and K. Kajita, *J. Phys. Soc. Jpn.*, **71**, 1832 (2002).
17. J. Merino and R.H. McKenzie, *Phys. Rev. Lett.*, **87**, 237002 (2001).
18. N. Drichko, M. Dressel, C.A. Kuntscher, A. Pashkin, A. Greco, J. Merino, and J. Schlueter, *Phys. Rev. B.,* **74**, 235121 (2006).
19. A.V. Dolbin, M.V. Khlistyuck, V.B. Esel'son, V.G. Gavrilko, N.A. Vinnikov, R.M. Basnukaeva, V.V. Danchuk, V.A. Konstantinov, Y. Nakazawa, *Low Temp. Phys.,* **42**, 788 (2016).
20. A.V. Dolbin, M.V. Khlistyuck, V.B. Esel'son, V.G. Gavrilko, N.A. Vinnikov, R.M. Basnukaeva, V.A. Konstantinov, Y. Nakazawa, *Low Temp. Phys.,* **43**, 1740 (2017).
21. Y. Nakazawa, A. Kawamoto, and K. Kanoda, Phys. Rev. B., **52**, 12890 (1995).
22. H. Taniguchi, Y. Nakazawa, K. Kanoda, Phys. Rev. B., **57**, 3623 (1998).
23. T. Nomoto *et al.* unpublished data.
24. S. Endo, Y. Watanabe, T. Sasaki et. al., *Synth. Met.*, **86,** 2013 (1997).